\documentclass[final,5p,times,twocolumn]{elsarticle}
\usepackage{amssymb}
\usepackage{graphicx}
\usepackage{amsmath,amssymb,amsfonts}
\usepackage{textcomp}
\usepackage{gensymb}
\usepackage[bookmarks=false]{hyperref}
\hypersetup{
		breaklinks=true,
    unicode=false,          
    pdftoolbar=true,        
    pdfmenubar=true,        
    pdffitwindow=false,     
    pdfstartview={FitH},    
    pdftitle={Taufour},    
    pdfauthor={Taufour},     
    pdfsubject={MnBi-Rh},   
    pdfcreator={Taufour},   
    pdfproducer={Taufour}, 
    pdfkeywords={keyword1} {key2} {key3}, 
    pdfnewwindow=true,      
    colorlinks=true,       	
    linkcolor=black,          
    citecolor=black,    	    
    filecolor=black,   		    
    urlcolor=blue            
}

\begin{document}

\title{A study of the physical properties of single crystalline Fe$_{5}$B$_{2}$P.}

\author[label1,label2]{Tej N Lamichhane}  \author[label2]{Valentin Taufour} \author[label2]{Srinivasa Thimmaiah} \author[label3]{David S. Parker} \author[label1,label2]{Sergey L. Bud'ko}  \author[label1,label2]{Paul C. Canfield}

\address[label1]{Department of Physics and Astronomy, Iowa State University, Ames, Iowa 50011, U.S.A.}

\address[label2]{Ames Laboratory, Iowa State University, Ames, Iowa 50011, U.S.A.}

\address[label3] {Materials Science and Technology Division, Oak Ridge National Laboratory, Oak Ridge, TN 37831}

\begin{abstract}
Single crystals of Fe$_{5}$B$_{2}$P were grown by self-flux growth technique. Structural and magnetic properties are studied. The Curie temperature of Fe$_{5}$B$_{2}$P is determined to be $655\pm2$~K. The saturation magnetization is determined to be $1.72$~${\mu}_{B}$/Fe at $2$~K. The temperature variation of the anisotropy constant $K_{1}$ is determined for the first time, reaching $\sim0.50$~MJ/m$^{3}$ at $2$~K, and it is comparable to that of hard ferrites. The saturation magnetization is found to be larger than the hard ferrites. The first principle calculations of saturation magnetization and anisotropy constant are found to be consistent with the experimental results.

\end{abstract}


\maketitle
\section{Introduction}

The existence of the ternary Fe$_{5}$B$_{2}$P phase was first reported in 1962\cite{Fruchart1962,Rundqvist1962ACS}. Both references reported the detailed structural information and the Curie temperature for the Fe$_{5}$B$_{2}$P phase. Its structural prototype is tetragonal Cr$_{5}$B$_{3}$ with the space group D$_{4h}^{ls} - I4/mcm$. The Curie temperature was reported to fall between $615$~K to $639$~K, depending upon the B content. In 1967, another study reported a Curie temperature of $628$~K and a saturation magnetization of $1.73$~${\mu}_{B}$/Fe~\cite{Blanc1967}. The Fe$_{5}$B$_{2}$P phase was also studied using M{\"o}ssbauer spectroscopy and X-ray diffraction in 1975~\cite{Lennart1975}. In addition to confirming the Curie temperature range as well as the average saturation magnetic moment per Fe atom, the M{\"o}ssbauer study identified the average moment contributed by each of the Fe lattice sites in the Fe$_{5}$B$_{2}$P unit cell. The Fe(2) (or 4c) sites contribute $2.2$~$\mu_B$/Fe. The Fe(1) (or 16l) sites contribute $1.6$~$\mu_B$/Fe. The average extrapolated moment of the both sites at $0$~K was reported to be $1.73$~$\mu_B$/Fe.   

Fe$_{5}$B$_{2}$P is specifically interesting as a possible high transition temperature, rare earth free, hard ferromagnetic material. Given that all prior work on Fe$_{5}$B$_{2}$P was made on polycrystalline samples, we developed a single crystal growth protocol, measured thermodynamic and transport properties of single crystalline samples, and determined the magnetic anisotropy of this material. The anisotropy constant $K_1$ is positive, indicating that the c axis is the easy axis of magnetization, and has a comparable size and temperature dependence as hard ferrites such as  SrFe$_{12}$O$_{19}$ and BaFe$_{12}$O$_{19}$.
\section{Experimental Details}

\subsection{Crystal growth}

As part of our effort to search for new, or poorly characterized ferromagnetic compounds, we have developed single crystal growth protocols for transition metal rich, chalcogenide and pnictide binary and ternary phases.  In a manner similar to some of our earlier transition metal - sulphur work, \cite{Xiaolin2012, Lin201541}  we started by confirming our ability to contain Fe-P binary melts in alumina crucibles sealed in amorphous silica ampules. As outlined by Canfield and Fisk \cite{CanfieldFisk} and Canfield, \cite{PCbeginerguide} sealed ampoules were decanted after slow cooling by use of a centrifuge. Crucibles with alumina filters \cite{PetrovicCanfield} were used to allow assessment and even reuse of the decanted liquid. For this experiment, a mixture of freshly ball milled iron powder and red phosphorous lumps were placed in an alumina crucible in an atomic ratio of Fe\,:\,P\,=\,0.83\,:\,0.17. A homogenous liquid exists at $1060$~\degree C (i.e. there was no crystal growth upon cooling from $1200$~\degree C to $1060$~\degree C and all of the material decanted). For similar temperature profiles, an initial melt of Fe$_{0.86}$P$_{0.14}$ lead to the growth of dendritic Fe whereas for initial melts of Fe$_{0.81}$P$_{0.19}$, Fe$_{0.79}$P$_{0.21}$ and Fe$_{0.77}$P$_{0.23}$ faceted Fe$_{3}$P was grown.  These data are all consistent with the binary phase diagram~\cite{ASMFeP} and indicate that the Fe-P binary melt does not have a significant partial pressure of phosphorous and does not react with alumina.

\begin{figure}[!htbp]
\begin{center}
\includegraphics[width=4.2cm]{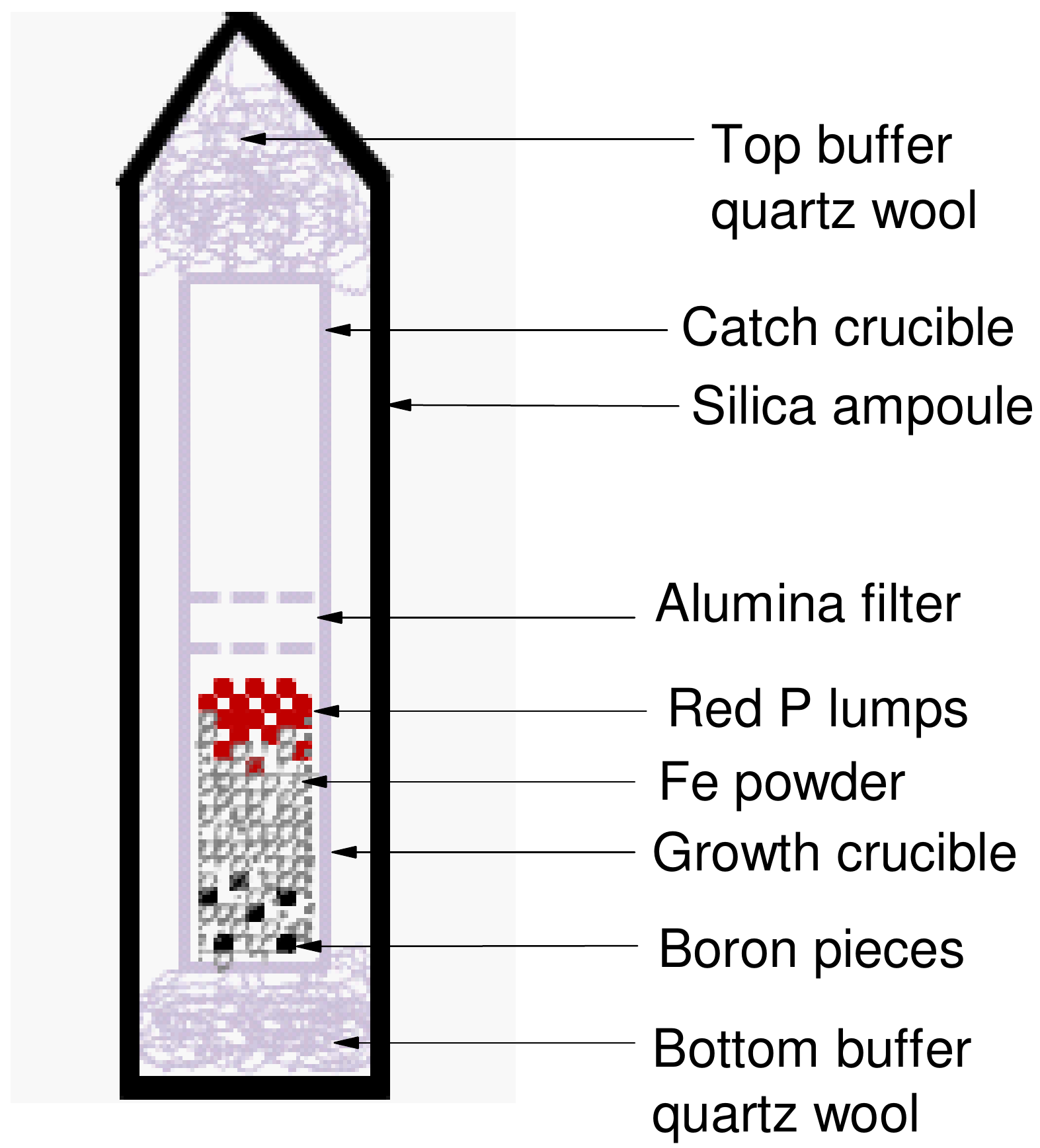}
\caption{\label{Growthpackup}A schematic assembly of the crystal growth ampoule.}
\end{center}
\end{figure}

\begin{figure}[!htbp]
\begin{center}
\includegraphics[width=4cm]{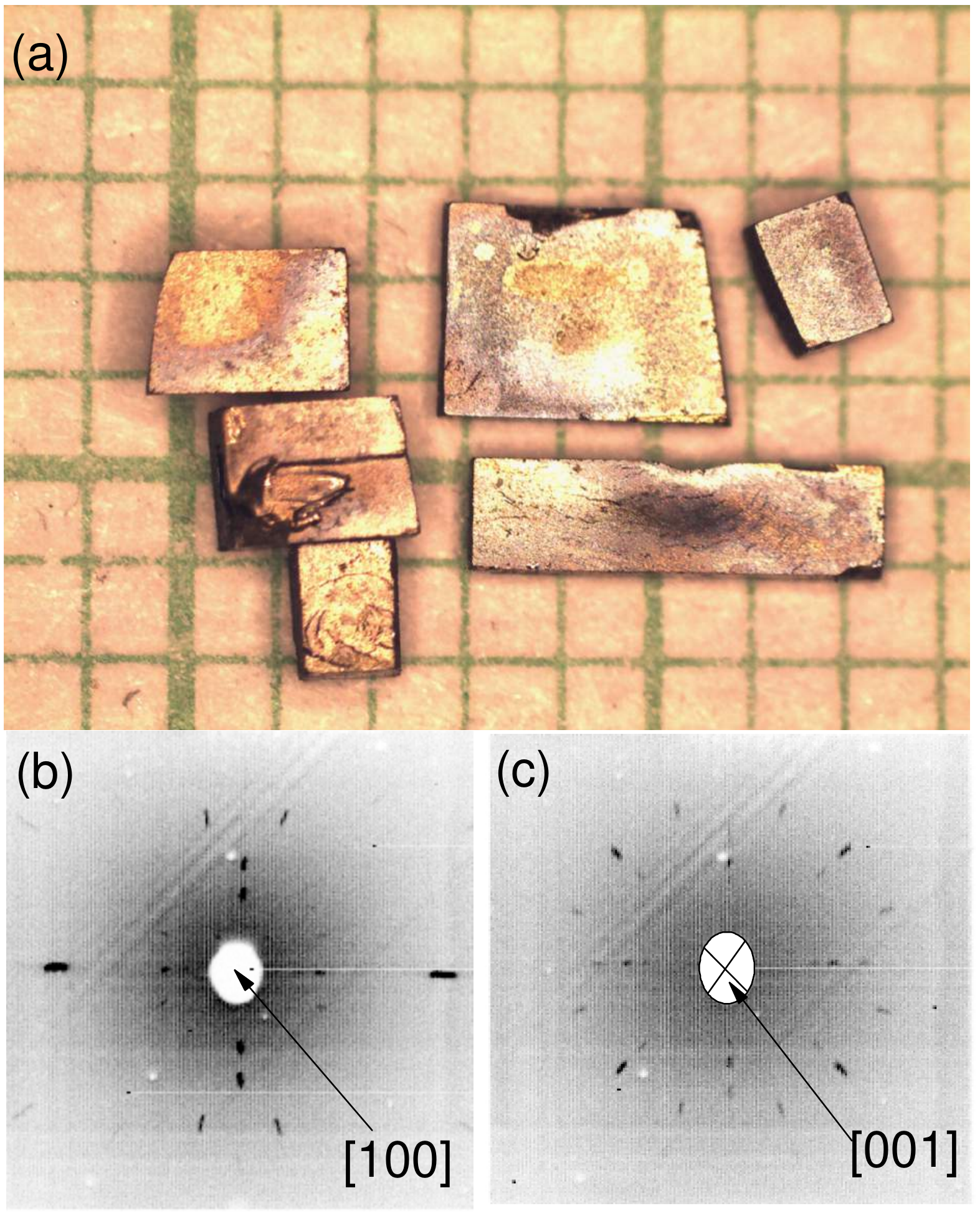}
\caption{\label{fig:crystalsabc}(a) The acid etched single crystals image of Fe$_{5}$B$_{2}$P (b) Laue pattern along the hard axis [100] and (c) Laue pattern along the easy axis [001] of magnetization.}
\end{center}
\end{figure}

After some optimization, an initial stoichiometry of Fe$_{72}$P$_{18}$B$_{10}$ was used to grow single phase Fe$_{5}$B$_{2}$P plates.  Ball milled Fe (Fe lumps obtained from Ames lab), red phosphorous lumps (Alfa Aesar, 99.999\% (metal basis)), and crystalline boron pieces (Alfa Aesar, 99.95\%) were placed in an alumina crucible / filter assembly, sealed in a partial pressure of Ar in an amorphous silica tube (as shown schematically in figure~\ref{Growthpackup}).  The ampoule was heated over 3 hours to $250$~\degree C, remained at $250$~\degree C for $3$~hours, heated to $1200$~\degree C over $12$~hours, held at $1200$~\degree C for $10$~hours, and then cooled to $1160$~\degree C over $75$~hours.  After cooling to $1160$~\degree C the ampoule was decanted using a centrifuge and plate like single crystals of Fe$_{5}$B$_{2}$P could be found on the growth side of the alumina filter. In order to confirm that the growth of crystals took place from a complete liquid, we decanted one growth at $1200$~\degree C, instead of cooling to $1160$~\degree C, and indeed found all of the material decanted. \\*
\indent
After growth, single crystals were cleaned by etching in a roughly 6 molar HCl solution. Figure~\ref{fig:crystalsabc}(a) shows a picture of the etched single crystals.

\subsection{Physical properties measurement}

The crystal structure and lattice parameters of Fe$_{5}$B$_{2}$P were determined with both single crystal and powder x-ray diffraction (XRD). The crystal structure of Fe$_{5}$B$_{2}$P was determined from single-crystal XRD data collected with the use of graphite monochromatized MoK$\alpha$ radiation ($\lambda=0.71073$~\AA) at room temperature  on a Bruker APEX2 diffractometer. Reflections were gathered by taking four sets of 360 frames with $0.5^\circ$ scans in $\omega$, with an exposure time of $25$~s per frame and the crystal-to-detector distance was $5$~cm. The measured intensities were corrected for Lorentz and polarization effects. The intensities were further corrected for absorption using the program SADABS, as implemented in Apex 2 package \citep{Bruker}.
 
 For powder XRD, etched single crystals of Fe$_{5}$B$_{2}$P were selected and finely powdered. The powder was evenly spread over the zero background single crystal silicon wafer sample holder with help of a thin film of Dow Corning high vacuum grease. The powder diffraction pattern was recorded with Rigaku Miniflex diffractrometer using copper $K_{\alpha}$ radiation source over 8.5 hours (at a rate of $3$~sec dwell time for per $0.01^\circ$ to cover the $2\theta$ value up to $100^\circ$).\\*
 
  To identify the crystallographic orientation of the single crystal plates, Laue diffraction patterns were obtained using a Multiwire Laboratories, Limited spectrometer. The resistivity data were measured in a four-probe configuration using a Quantum Design Magnetic Property Measurement System (MPMS) for temperature control and the external device control option to interface with a Linear Research, Inc. ac (20mA, 16 Hz) resistance bridge (LR 700).\\*

The sample preparation for magnetization measurements is a major step in a magnetic anisotropy study. The etched crystal was cut into a rectangular prismatic shape and the dimensions were determined with a digital Vernier caliper. 

	Temperature and field dependent magnetization was measured using the MPMS up to room temperature and a Quantum Design Versalab Vibration Sample Magnetometer (VSM) with an oven option for higher temperature ($T<1000$~K). In MPMS, plastic straw was used to align the sample in desired directions. The sample was glued to the VSM sample heater stick with Zircar cement obtained from ZIRCAR Ceramics Inc.. While gluing, the sample was pushed into the thin layer of Zircar paste spread on the heater stick to ensure a good thermal contact with the heater stick. When the sample was firmly aligned with the desired direction it was covered with Zircar cement uniformly. Finally, the VSM heater stick, with the sample glued on it, was covered with a copper foil so as to (i) better control the heat radiation in the sample chamber, (ii) maintain a uniform temperature inside the wrapped foil (due to its good thermal conductivity), and (iii) further secure the sample throughout the measurement.

	In the VSM, both zero field cooled (ZFC) as well as field cooled (FC) magnetizations were measured and found to be almost overlapping. The difference between the measured data in the VSM and the MPMS was found to be less than $3$\% at $300$~K (i.e. at the point of data overlap). We normalized the magnetization data from the MPMS with FC VSM data to get a smooth curve for the corresponding applied field.
 
\subsection{Determination of demagnetization factor for transition temperature and anisotryopy constant measurement}
\begin{figure}[!htbp]
\begin{center}
\includegraphics[width=7cm]{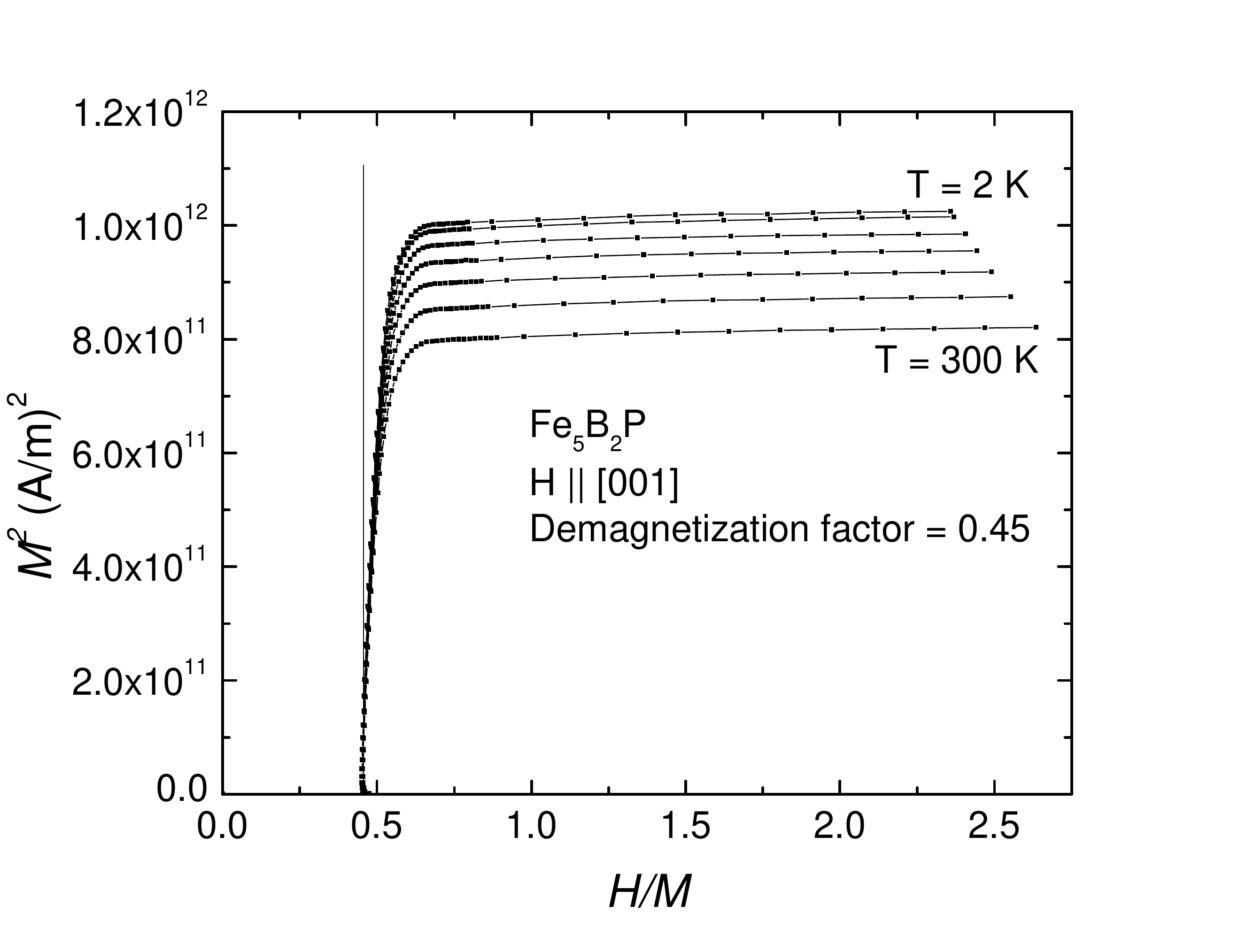}
\caption{\label{DemagfactorMPMS}An analysis of M(H) isotherms taken at $T=2$, $50$, $100$, $150$, $200$, $250$, $300$~K  and plotted as M$^{2}$ versus $H/M$ to determine the demagnetization factor. The $H/M$ axis intercept at $0.45$ is the experimental demagnetization factor for easy axis of magnetization.}
\end{center}
\end{figure}

The demagnetization factors along different directions were determined by using a formula developed by Aharoni \cite{Aharoni1998}. The calculated demagnetization factors for the field along a, b, and c axes were determined to be 0.21, 0.29, and 0.50 respectively. To verify that demagnetization factors were accurate, we prepared a  M$^{2}$ versus $\frac{H}{M}$ plot for the lower temperature M(H) data along the easy axis of magnetization as shown in figure~\ref{DemagfactorMPMS}. The X-intercept gives the directly measured experimental value of the demagnetization factor along the easy magnetization axis \cite{Arrott1957}. In figure~\ref{DemagfactorMPMS}, we can clearly see that all the M$^{2}$ curves are overlapping near the M$^{2}$ axis indicating that the demagnetization factor along the easy magnetization direction does not depend on temperature. A value of demagnetization factor of 0.45 was determined along the c axis which is not that different from the value of $0.50$ inferred from the sample dimensions. Based on this result, we readjusted the two hard axes demagnetization factors in proportion such that the total sum of all 3 of them is 1. The experimentally readjusted values for demagnetization factors were $0.231$, $0.319$ and $0.45$ along a, b and c axes respectively. With the benefit of fourfold symmetry of Fe$_{5}$B$_{2}$P unit cell perpendicular to its c axis, magnetization was measured along a and c axes and corresponding demagnetization factors were used to calculate the corrected internal magnetic field ($H_{\textrm{int}}$). Here $H_{\textrm{int}}= H_{\textrm{applied}}-NM$, where $N$ is the demagnetization factor and $M$ is the magnetization.
 
\section{Results and discussion}

\begin{figure}[!htbp]
\begin{center}
\includegraphics[width=1.0\linewidth]{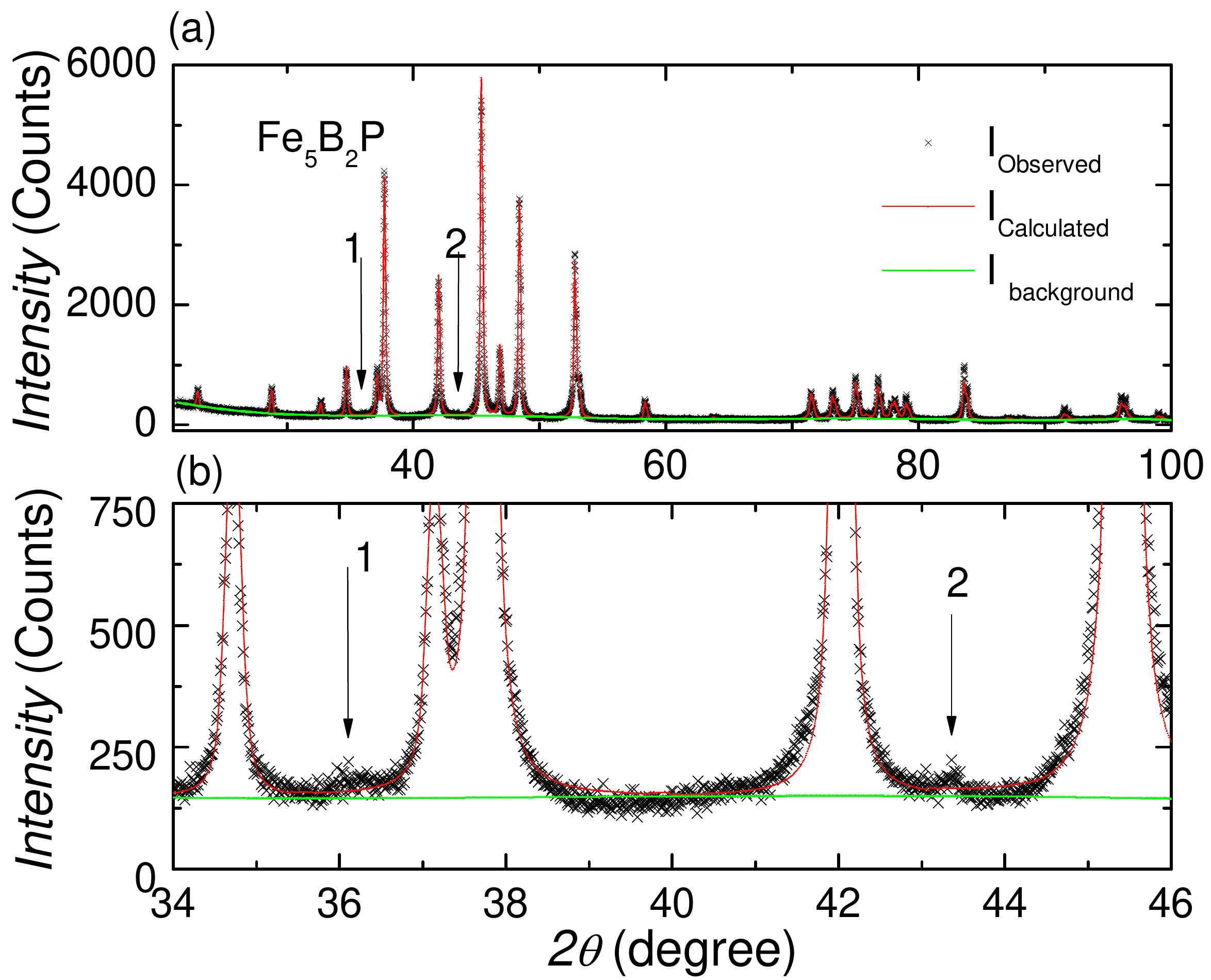}
\caption{\label{xrayab}(a) Powder x-ray diffraction pattern of Fe$_{5}$B$_{2}$P (b) Enlarged powder X-ray diffraction pattern in between 2$\theta$ value of $34^\circ$ to $46^\circ$ to show the weak impurity peaks 1 and 2.}
\end{center}
\end{figure}
\subsection{Lattice parameters determination}


The structure solution and refinement for single crystal data was carried out using SHELXTL program package \cite{SHELXTL}.  The final stage of refinement was performed using anisotropic displacement parameters for all the atoms. The refined composition was Fe$_{5}$B$_{2.12}$P$_{0.88(1)}$ with residual $R1 = 2.3 \%$ (all data). The off-stoichiometry was due to partial replacement of phosphorus by boron on 4a site in the structure. According to  the previous report \citep{Lennart1975} Fe$_{5}$B$_{2}$P shows a considerable phase width. 
All the details about the atomic positions, site occupancy factors, and displacement parameters for crystal of Fe$_{5}$B$_{2.12}$P$_{0.88(1)}$ are given in tables~\ref{tbl:crystaldata}, and~\ref{tbl:atomcoord}.\\*

\begin{table}[!htbp]
\begin{center}
\caption{\label{tbl:crystaldata}Crystal data and structure refinement for Fe$_{5}$B$_{2}$P.}
\begin{tabular}{|l|l|}
\hline
Empirical formula & Fe$_{5}$B$_{2.12}$P$_{0.88}$ \\
Formula weight & 329.42 \\
Temperature & $293(2)$~K \\
Wavelength & $0.71073$~\AA \\
Crystal system, space group & Tetragonal,  \textit{I}4/\textit{mcm} \\
Unit cell dimensions & a=5.485(3)~\AA \\
 & b = 5.485(3)~\AA \\
 & c = 10.348(6)~\AA \\
Volume & 311.3(4) $10^3$ \AA$^3$ \\
Z, Calculated density & 4,  7.029 g/$cm^3$ \\
Absorption coefficient & 22.905 mm$^{-1}$ \\
F(000) & 615 \\
Crystal size & 0.01 x 0.05 x 0.08 mm$^3$ \\
$\theta$ range ($^\circ$) & 3.938 to 31.246 \\
Limiting indices & $-7\leq h \leq7$\\
& $-7\leq k\leq 7$\\
& $-14\leq l \leq 14$ \\
Reflections collected & 2113 \\
Independent reflections & 152 [R(int) = 0.0433] \\
Completeness to $\theta = 25.242^\circ$ & $100.00$\% \\
Absorption correction & multi-scan, empirical \\
Refinement method & Full-matrix least-squares \\
& on F$^2$ \\
Data / restraints / parameters & 152 / 0 / 17 \\
Goodness-of-fit on $F^2$ & 1.101 \\
Final R indices [I$>2\sigma$(I)] & $R1 = 0.0140$, $wR2 = 0.0289$ \\
R indices (all data) & $R1 = 0.0180$, $wR2 = 0.0299$\\
Extinction coefficient & 0.0243(3) \\
Largest diff. peak and hole & 0.485 and -0.474 e.\AA$^{-3}$\\ \hline
\end{tabular}
\end{center}
\end{table}

\begin{table}[!htbp]
\begin{center}
\caption{\label{tbl:atomcoord}Atomic coordinates and equivalent isotropic displacement parameters (A$^2$) for Fe$_{5}$B$_{2}$P. U(eq) is defined as one third of the trace of the orthogonalized U$_{ij}$ tensor.}
\begin{tabular}{| p{0.6cm} | p{0.85cm} | p{1.2cm} | p{1.2cm} | p{1.2cm} |p{1cm}|}
\hline
atom & Occ & x & y & z & U$_{eq}$ \\ \hline
Fe1 & 1 & 0.0000 & 0.0000 & 0.0000 & 0.005(1) \\ \hline
Fe2 & 1 & 0.1701(1) & 0.6701(1) &0.1403(1) & 0.005(1) \\ \hline
P3 & 0.88(1) & 0.0000 &  0.0000 & 0.2500 & 0.004(1) \\ \hline
B3 & 0.12(1) &  0.0000 & 0.0000 & 0.2500 & 0.004(1) \\ \hline
B4 & 1 & 0.6175(2) & 0.1175(2) &  0.0000 & 0.005(1) \\ \hline
\end{tabular}
\end{center}
\end{table}


Fe$_{5}$B$_{2}$P has a tetragonal unit cell with lattice constants a\,=\,5.485\,(3)\,~\AA and $c = 10.348(3)$~\AA. While analysing the powder pattern,  the CIF file obtained from single crystal data was used and the powder pattern was fitted with Rietveld analysis using GSAS EXPGUI software package \cite{GSAS, EXPGUI}. During the Rietveld analysis, Fe sites were supposed to be fully occupied and thermally rigid whereas P and B occupation number were released between each other with help of constraints. Finally, a well fitted powder diffraction pattern with R$_{p}$ = 0.0828 was obtained as shown in figure~\ref{xrayab}(a). The lattice parameters from this measurement are in close agreement (less than 0.2 \% deviation) with our single crystal data as well as previously reported data \cite{Rundqvist1962ACS, Lennart1975}. A final stoichiometry of the powder sample was determined to be Fe$_{5}$B$_{2.11}$P$_{0.89}$ . This stoichiometry is in agreement with our single crystal XRD composition of Fe$_{5}$B$_{2.12}$P$_{0.88(1)}$. Two tiny unidentified peaks denoted by $1$ and $2$ in figure~\ref{xrayab} were noticed in all batches of Fe$_{5}$B$_{2}$P measured. A possible origin for these peaks is an excess amount of unreacted boron trapped in crystal. We suspected boron because it has many overlapping diffraction peaks with Fe$_{5}$B$_{2}$P as well as with these two tiny humps denoted by 1 and 2 and enlarged in figure~\ref{xrayab}(b).
 
\subsection{Identification of crystallographic orientation}

The Laue diffraction pattern shown in  figure~\ref{fig:crystalsabc}(b) was obtained with the X-ray beam parallel to the plane of the plate [100]. The Laue pattern shown in figure~\ref{fig:crystalsabc}(c) was obtained with the X-ray beam perpendicular to the plane of the plate [001]. The obtained Laue diffraction patterns were analysed with the OrientExpress analysis software~\cite{OrientExpress}. The Laue pattern analysis revealed that the crystals facets were grown along [100], [010], and [001] directions.

\subsection{Resistivity measurement}

\begin{figure}[!htbp]
\begin{center}
\includegraphics[width=7cm]{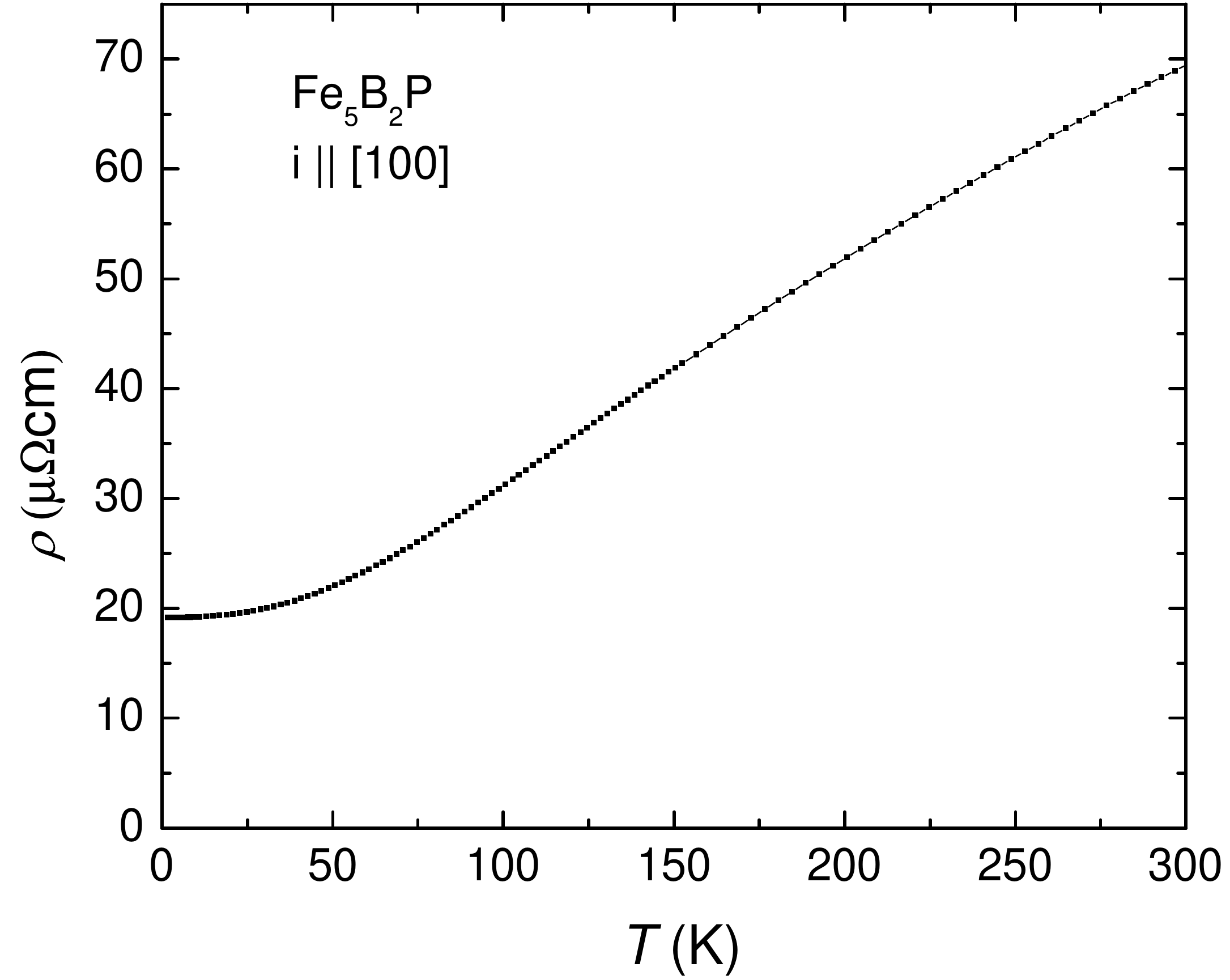}
\caption{\label{fig:res}Resistivity of Fe$_{5}$B$_{2}$P below the room temperature with an excitation current being parallel to [100] direction.}
\end{center}
\end{figure}

The resistivity data of a single crystalline sample helps to test its quality. The resistivity of Fe$_{5}$B$_{2}$P with current parallel to [100] is measured from room temperature down to $2$~K; the resistivity is found to be metallic in nature as shown in figure~\ref{fig:res}. The residual resistivity ratio ($RRR=\rho(300\textrm{~K})/\rho(2\textrm{~K})$) of the Fe$_{5}$B$_{2}$P sample is estimated to be nearly $3.6$. The residual resistivity $\rho{(2\textrm{~K})}$ is roughly $20$~$\mu\Omega$cm. These values are consistent with some residual disorder in the sample (e.g. the P an B site disorder on the P$_{3}$ site).

\begin{figure}[!htbp]
\begin{center}
\includegraphics[width=7cm]{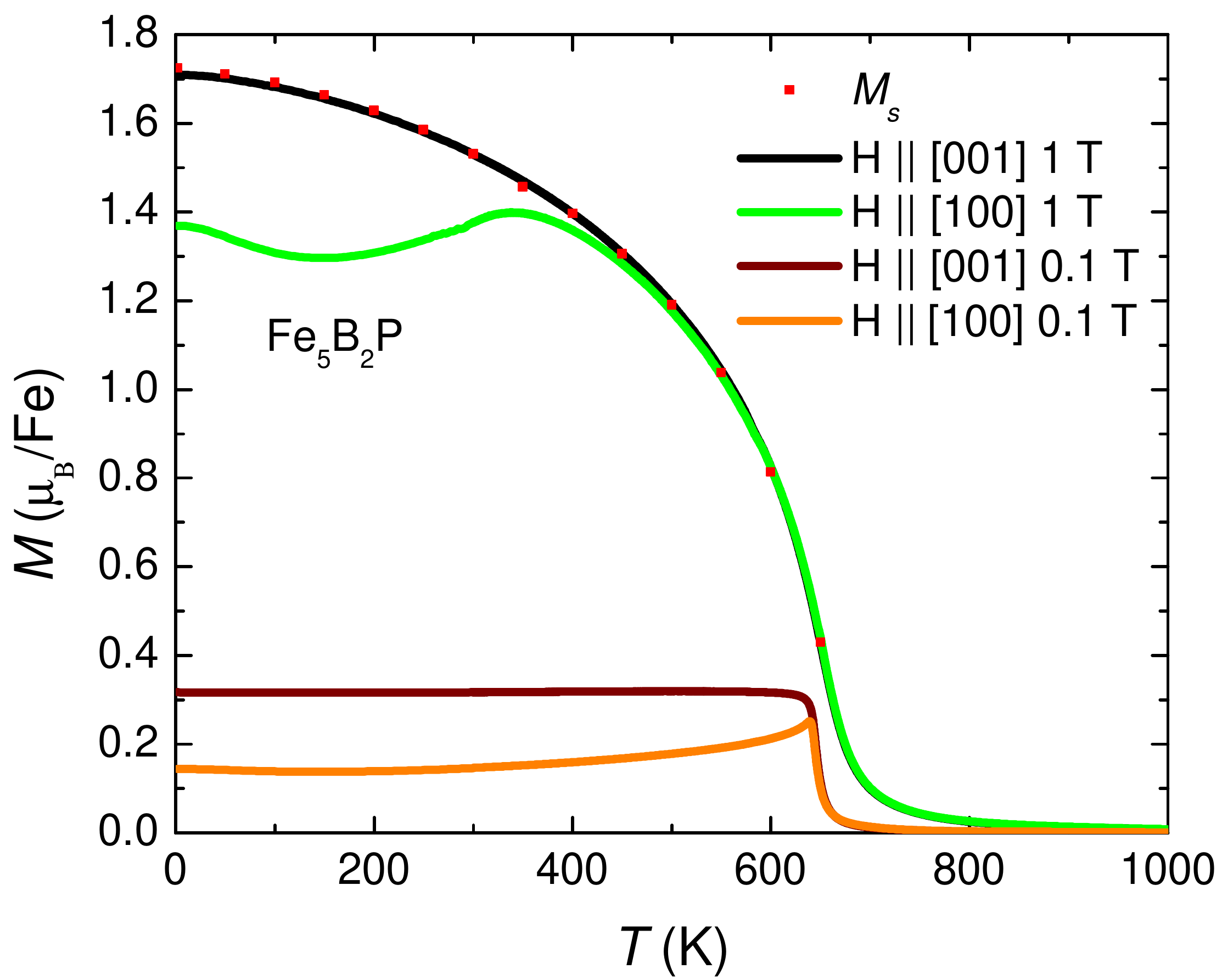}
\caption{\label{CombinedMTMS}Temperature dependent magnetization of Fe$_{5}$B$_{2}$P at various magnetizing field along [100] and [001] directions and saturation magnetization, $M_{s}$, inferred from M(H) isotherms.}
\end{center}
\end{figure}

\subsection{Measurement of magnetization and saturation magnetization}

\begin{figure}[!htbp]
\begin{center}
\includegraphics[width=7cm]{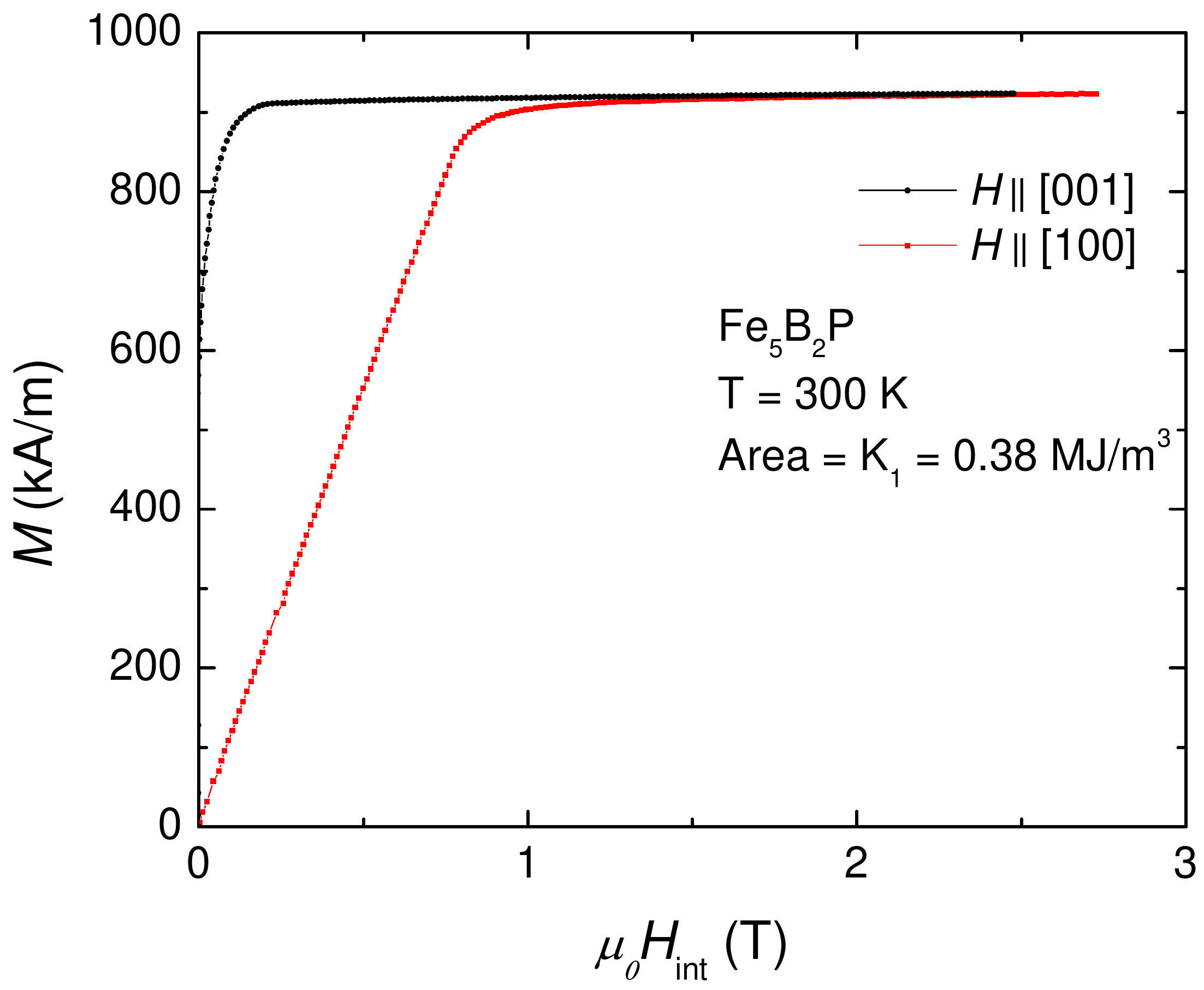}
\caption{\label{fig:K1300K}An example of determination of K$_{1}$ at $300$~K with an appropriate axes units so as to obtain the area in terms of MJ/$m^3$ unit.}
\end{center}
\end{figure}

The temperature dependent magnetization of Fe$_{5}$B$_{2}$P along the easy [001] and hard [100] axes in various magnetizing field strengths in terms of ${\mu}_{B}$/Fe are reported in figure~\ref{CombinedMTMS}. As the field strength increases, the moment increases and saturates. The saturating field for the easy axis was found to be nearly $0.8$~T and $M(T)$ data in an applied field of $1$~T is the same as saturation magnetization obtained from $M(H)$ isotherms data (not shown here). The magnetization along the hard axis has not reached saturation in an external field of $1$~T as shown in the green curve of figure~\ref{CombinedMTMS}. Below the Curie temperature, the $M(T)$ curves at low field are almost constant because the applied field is less than the demagnetizing field. The saturation magnetization ($M_s$) data were determined from the Y-intercept of linear fit of $M(H)$ isotherms plateau (as shown in figure~\ref{fig:K1300K})  at $2$~K, $50$~K and in an interval of $50$~K up to $650$~K. The saturation magnetization at $2$~K was found to be $1.72$~$\mu_B$/Fe which is very close to the previously reported value of $1.73$~$\mu_B$/Fe \cite{Blanc1967, Lennart1975}.           
      
\begin{figure}[!htbp]
\begin{center}
\includegraphics[width=7cm]{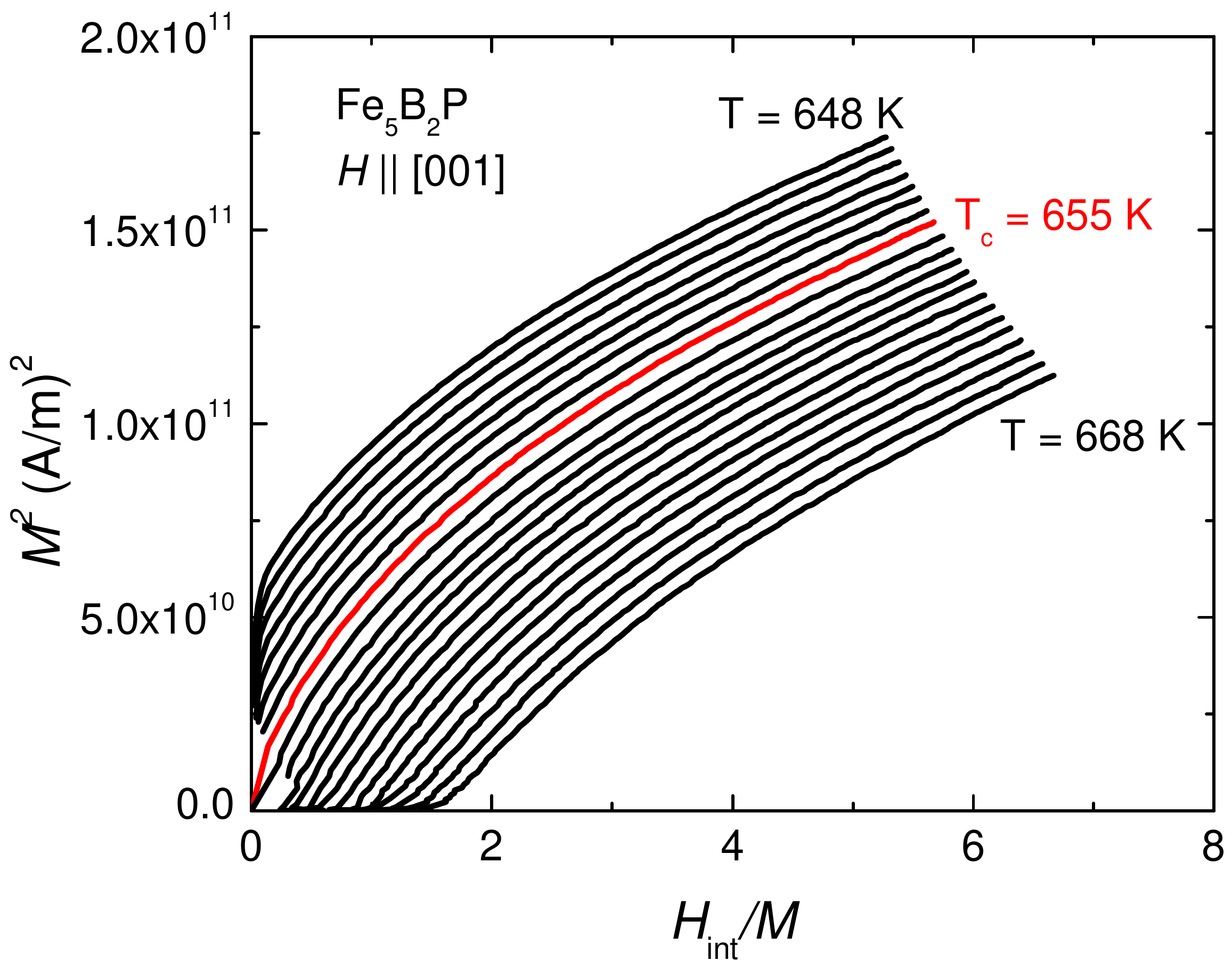}
\caption{\label{fig:Arrott}The Arrott plot of Fe$_{5}$B$_{2}$P. Here M(H) isotherms were measured for the prismatic sample from $648$~K to $668$~K with a spacing of $1$~K. Internal magnetic field ($H_{\textrm{int}}$) was determined with an experimentally measured demagnetization factor ($0.45$) for the easy axis of magnetization. The temperature corresponding to $M^2$ versus $H_{\textrm{int}}/M$ isotherm passing through origin gives the Curie temperature. The Curie temperature is determined to be $655\pm2$~K.}
\end{center}
\end{figure}

\subsection{Determination of transition temperature}
A M$^{2}$ versus $H_{\textrm{int}}/M$ Arrott plot was used to determine the Curie temperature. The Curie temperature corresponds to Arrott plot curve that passes through the origin.
For our sample, the Arrott curve corresponding to $655$~K is passing through the origin. Hence the Curie temperature of Fe$_{5}$B$_{2}$P is determined to be $655\pm2$~K, where the error includes an instrumental uncertainty of $\pm1$~K inherent to  such a high temperature measurement in a VSM and a reproducibility error due to sample gluing process on the heater stick resulting in a variation in thermal coupling. This Curie temperature is a little bit higher than the previously reported window of $615$~K to $639$~K \cite{Blanc1967,Lennart1975}.\\*
\indent
	 To make sure our measurement was correct, magnetization of a piece of a nickel wire obtained from the Alfa Aesar company ($99.98$~\% metal basis) was measured with the same VSM heater stick.  Using the criterion from reference \cite{Arrott1967}, the Curie temperature of nickel sample was determined to be $625\pm2$~K which is in agreement with VSM Tech Note \cite{VSMappnote}. The Curie temperature of nickel is reported to fall between 626 to $633$~K \cite{Legendre2011}. These results confirm the accuracy of measured Curie temperature.    

\subsection{Determination of anisotropy constant K$_{1}$}

The anisotropy constant K$_{1}$ is the measure of anisotropy energy density and strongly depends on the unit cell symmetry and temperature. One of the conceptually simplest methods of measuring the anisotropy constant of an uniaxial system is to determine the area between the easy and hard axes  M(H) isotherms \cite{Bozorth1936}.  Here we measured both the easy and hard axes isothermal M(H) curves starting from $2$~K. Then we measured M(H) curves from $50$~K to $800$~K in $50$~K intervals. A typical example of determination of the anisotropy constant by measuring the anisotropy area between two magnetization curves is shown in figure~\ref{fig:K1300K}.

\begin{figure}[!htbp]          
\begin{center}
\includegraphics[width=7cm]{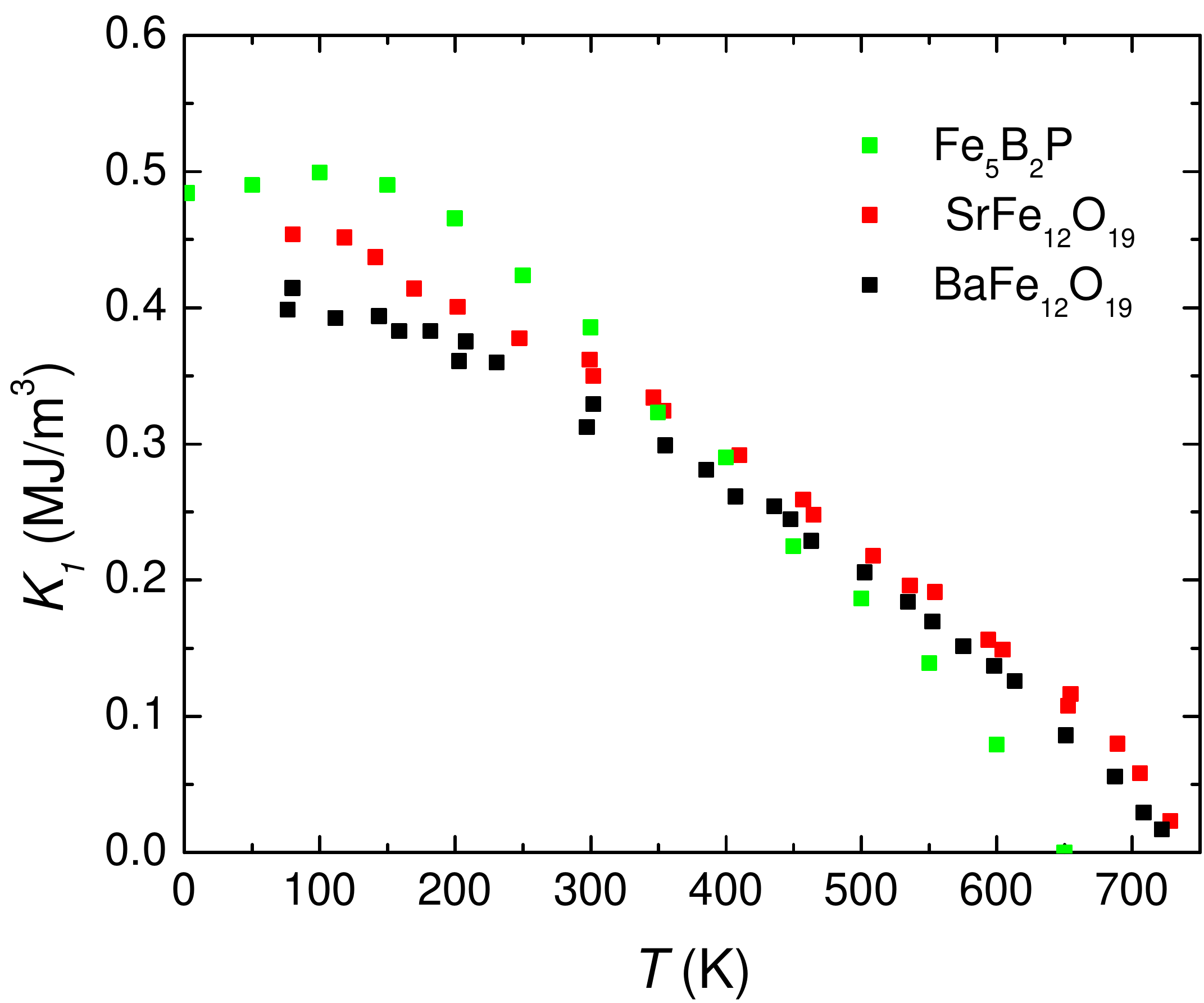}
\caption{\label{fig:K1Comparison}Temperature variation of the anisotropy constant K$_{1}$ of Fe$_{5}$B$_{2}$P and comparison with SrFe$_{12}$O$_{19}$ and BaFe$_{12}$O$_{19}$ from ref.~\cite{Shirk1969}.}
\end{center}
\end{figure}

The temperature dependence of $K_1$ is shown in Fig.~\ref{fig:K1Comparison}. $K_1$ is positive, indicating that the c axis is the easy axis of magnetization, in agreement with a previous calculation using a simple point charge model~\cite{Lennart1975}.

\begin{figure}[!htbp]
\begin{center}
\includegraphics[width=7cm]{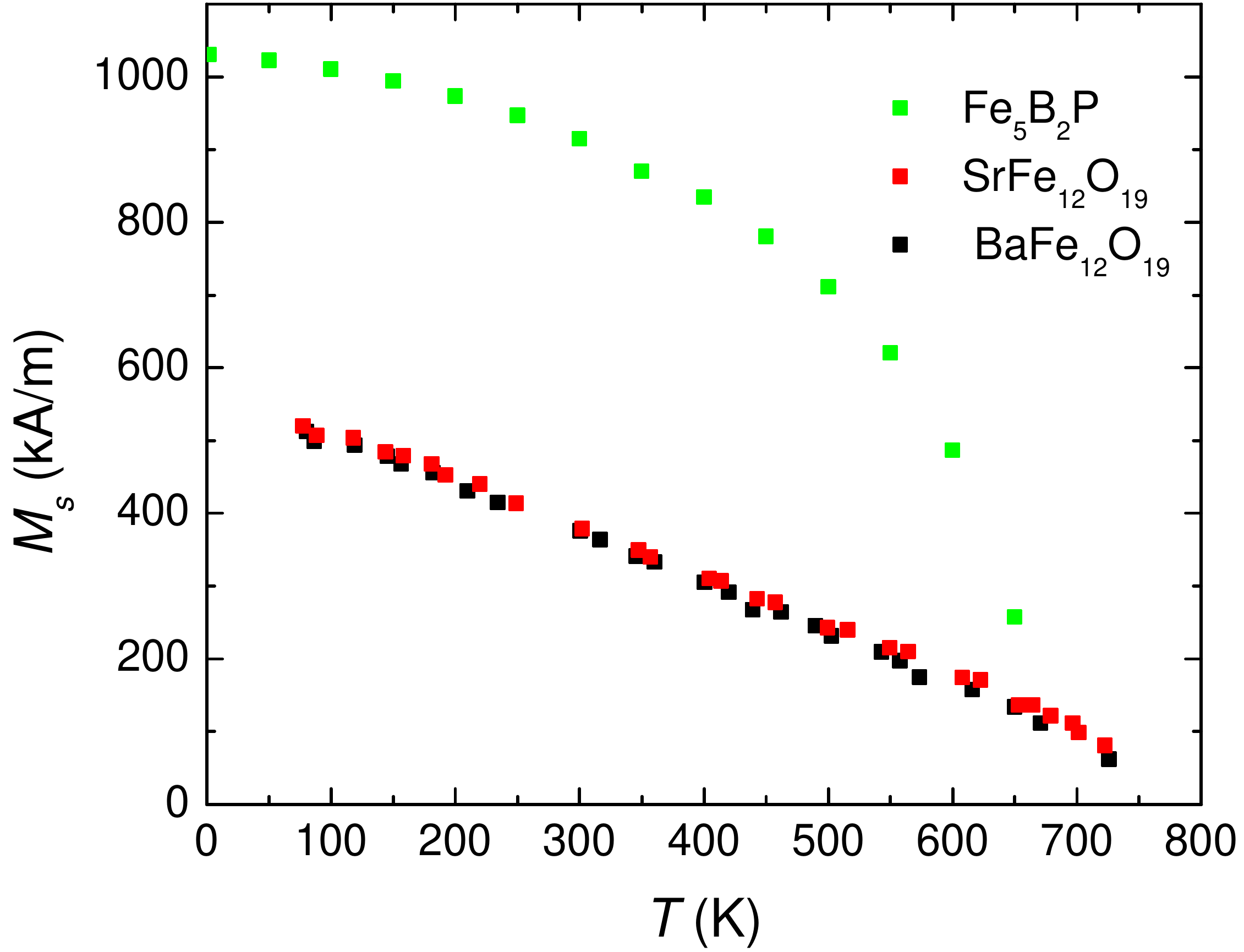}
\caption{\label{MsComparisonperFe}Temperature variation of the saturation magnetization M$_{s}$ of Fe$_{5}$B$_{2}$P and comparison with SrFe$_{12}$O$_{19}$ and BaFe$_{12}$O$_{19}$ from ref.~\cite{Shirk1969}.}
\end{center}
\end{figure}
 
We have compared the anisotropy constant $K_{1}$ and saturation magnetization $M_{s}$ of Fe$_{5}$B$_{2}$P with the single crystal data of hard ferrites SrFe$_{12}$O$_{19}$ and BaFe$_{12}$O$_{19}$~\cite{Shirk1969}. The anisotropy constant of Fe$_{5}$B$_{2}$P is greater at low temperature than either of these ferrites. The Fe$_{5}$B$_{2}$P sample becomes comparable to that of SrFe$_{12}$O$_{19}$ at room temperature but decreases slightly faster than both BaFe$_{12}$O$_{19}$ and SrFe$_{12}$O$_{19}$ above room temperature as shown in figure~\ref{fig:K1Comparison}. However, the nature of variation of the saturation magnetization of Fe$_{5}$B$_{2}$P is different than the hard ferrites as shown in figure~\ref{MsComparisonperFe}. Both of the ferrites show a roughly linear decrease of saturation magnetization with increasing temperature whereas the saturation magnetization of Fe$_{5}$B$_{2}$P is found to be significantly non-linear in T and also significantly larger for $T<600$~K. The saturation magnetization of Fe$_5$B$_2$P is $1.53$~$\mu_B$/Fe at $300$~K, which is already larger than the value of $1.16$~$\mu_B$/Fe in SrFe$_{12}$O$_{19}$. Since Fe$_5$B$_2$P contains less non-magnetic elements, the volume magnetization of $915$~kA/m for Fe$_5$B$_2$P is more than twice that of SrFe$_{12}$O$_{19}$ ($377$~kA/m~\cite{Shirk1969}).

\section{First principles calculations}

In an effort to understand the observed magnetic properties - in particular, the saturation magnetization and magnetocrystalline anisotropy - of Fe$_{5}$B$_{2}$P, we have performed first principles calculations using the augmented plane-wave density functional theory code WIEN2K~\cite{PBlaha2001} within the generalized gradient approximation (GGA) of Perdew, Burke and Ernzerhof~\cite{Perdew1996PRL}. Sphere radii of $1.77$, $2.15$ and $2.06$ Bohr were used for B, Fe and P, respectively, and an RK$_{max}$ of $7.0$ was employed, where RK$_{max}$ is the product of the smallest sphere radius and the largest plane-wave expansion wave vector. The experimental lattice parameters from Ref.~\cite{Rundqvist1962ACS} were used and all internal coordinates were relaxed. All calculations, the internal coordinate relaxation excepted, employed spin-orbit coupling and a total of approximately $10,000$~$k$-points in the full Brillouin zone were used for the calculation of magnetic anisotropy.  For these calculations we computed total energies for the magnetic moments parallel to (100) and (001) and computed the anisotropy as the difference in these energies.

\begin{figure}[!htbp]
\begin{center}
\includegraphics[width=7cm]{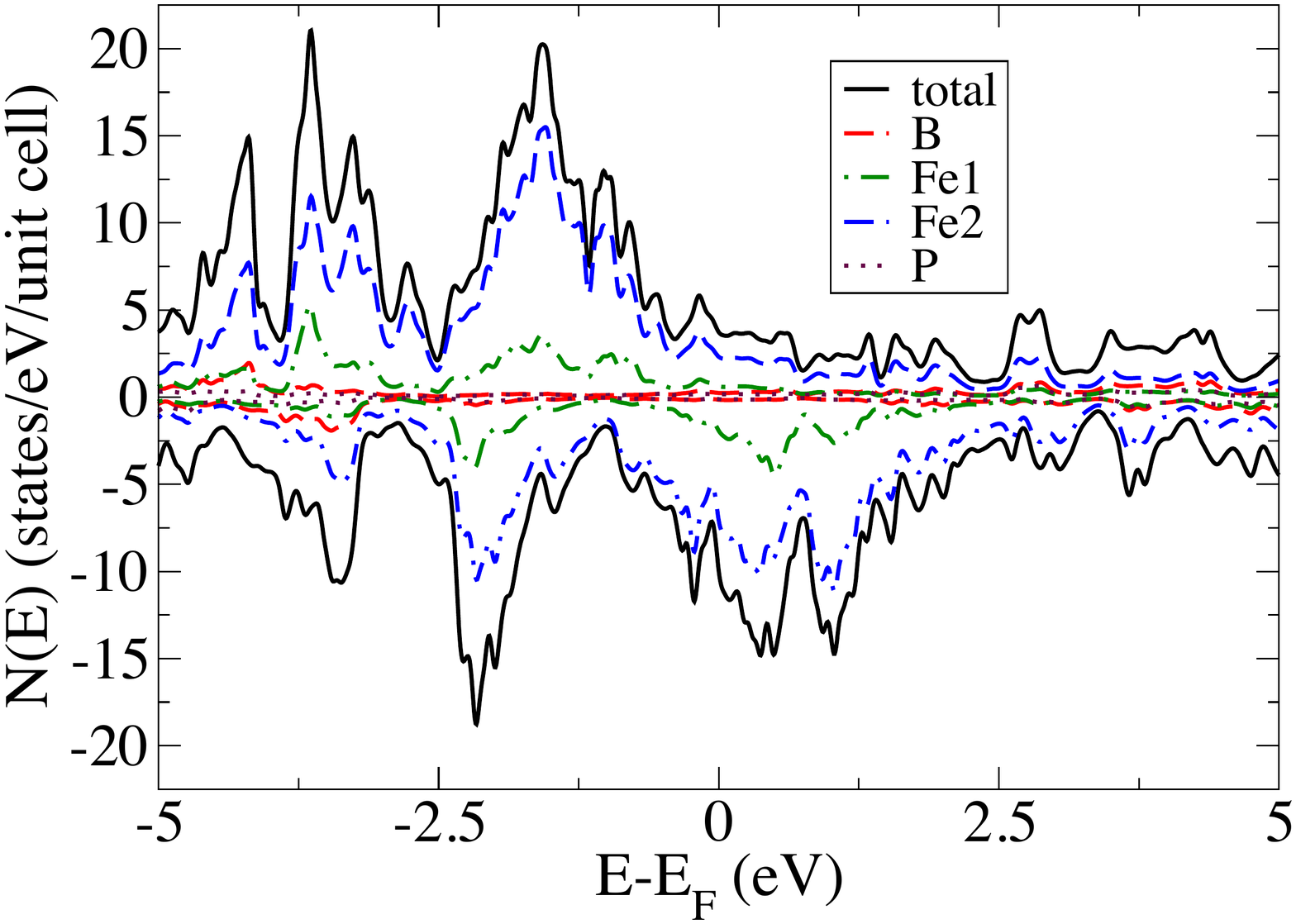}
\caption{\label{DOS}The calculated density-of-states of Fe5PB$_2$}
\end{center}
\end{figure}

As in the experimental work we find a strong ferromagnetic behavior in Fe$_{5}$B$_{2}$P, with a saturation magnetic moment of $1.79$~$\mu_{B}$~per~Fe, which includes an average orbital moment of approximately $0.03$~$\mu_{B}$ for each of the Fe sites.  We present the calculated density-of-states in Fig.~\ref{DOS}. The theoretical saturation value is in good agreement with the experimental $2$~K moment of $1.72$~$\mu_{B}$ per Fe.  Small negative moments of $-0.1$~$\mu_{B}$ and $-0.06$~$\mu_{B}$ per atom are found for the B and P atoms respectively, while the spin moments for the three distinct inequivalent Fe atoms are $1.79$, $1.79$ and $2.08$~$\mu_{B}$.  These are significantly smaller than the values for bcc Fe (approximately $2.2$~$\mu_{B}$) and for the Fe atoms in hexagonal Fe$_3$Sn (approximately $2.4$~$\mu_{B}$/Fe~\cite{Sales2014SR}), limiting the potential performance as a hard magnetic material. Figure~\ref{DOS} displays the reason for this, with the spin-minority DOS substantially larger than the spin-majority DOS around the Fermi energy, reducing the moment.

As in the experiment, the calculated easy axis of the magnetization is the $c$-axis, with a $T=0$ value for the first anisotropy constant K$_{1}$ of $0.46$~MJ/m$^{3}$, which is in excellent agreement with the experimental value of $0.48$~MJ/m$^{3}$.  On a per-Fe basis this is $44$~$\mu$eV.  This is much larger than the $\sim 1$~$\mu$eV value for bcc Fe~\cite{Daalderop1990PRB}, as might be expected given the non-cubic symmetry of Fe$_{5}$B$_{2}$P, but is roughly consistent with the $60$~$\mu$eV/Co value for hcp Co.  This is again indicative of the usual requirement of an anisotropic crystal structure for significant magnetic anisotropy.  Regarding the structure itself, in the perfectly ordered Fe$_{5}$B$_{2}$P structure, eight of the ten Fe atoms have three boron nearest neighbors and two next-nearest phosphorus neighbors, with the other two Fe atoms having four boron nearest neighbors.  There are no nearest-neighbor Fe-Fe bonds, which is perhaps surprising in a structure which is over 60 atomic percent Fe.  One may suppose that a variegated bonding configuration, with a range of atoms bonding with the Fe atoms, might be favorable for the magnitude, though not necessarily the {\it sign} (i.e. axial or planar) of the anisotropy, but this is apparently not realized in this material. In any case, the theoretical calculations are generally quite consistent with the results of the experimental work performed.

\section{Conclusions}

Single crystals of Fe$_{5}$B$_{2}$P were grown using a self flux growth method within a $40$~\degree C window of cooling. The Curie temperature of Fe$_{5}$B$_{2}$P was determined to be $655\pm2$~K. The saturation magnetization was determined to be $1.72$~${\mu}_{B}$/Fe at $2$~K. The temperature variation of the anisotropy constant $K_{1}$ was determined for the first time, reaching $\sim0.50$~MJ/m$^{3}$ at $2$~K, and found to be comparable to that of hard ferrites. The saturation magnetization, in unit of kA/m is found to be larger than the hard ferrites. The first principle calculation values of saturation magnetization and anisotropy constant using augmented plane-wave density functional theory code were found to be consistent with experimental work.
\section{Acknowledgement}
We thank T. Kong, U. Kaluarachchi, K. Dennis, and A. Sapkota for useful discussion. The research was supported by the Critical Material Institute, an Energy Innovation Hub funded by U.S. Department of Energy, Office of Energy Efficiency and Renewal Energy, Advanced Manufacturing Office. This work was also supported by the Office of Basic Energy Sciences, Materials Sciences Division, U.S. DOE. The first principle calculation of this work was performed in Oak Ridge National Laboratory.

\bibliographystyle{elsarticle-num}
\bibliography{TejResearch}

\end{document}